\RequirePackage{fix-cm}
\documentclass[smallextended]{svjour3}       
\smartqed  
\usepackage{amsmath}
\usepackage{graphics}
\usepackage{graphicx}
\usepackage{amssymb}
\usepackage{braket}
\usepackage{epstopdf}
\begin{document}
	\title{Quantum gate synthesis by small perturbation of a free particle in a box with electric field 
	}
	
	\titlerunning{}        
	
	\author{Kumar Gautam$^{1^*}$}
	
	\authorrunning{K.Gautam} 

	\institute{$^{1^*}$Quantum Computing lab, Quantum Research And Centre of Excellence, India\\ 
		\\ \email{Corresponding Author$^{1^*}$ : kgnsit@gmail.com}
		\and
	 }
	

	\maketitle

\begin{abstract}
\small{A quantum unitary gate is realized in this paper by perturbing a free charged particle in a one-dimensional box with a time- and position-varying electric field. The perturbed Hamiltonian is composed of a free particle Hamiltonian plus a perturbing electric potential such that the Schr$\ddot{o}$dinger evolution in time $T$, the unitary evolution operator of the unperturbed system after truncation to a finite number of energy levels, approximates a given unitary gate such as the quantum Fourier transform gate. The idea is to truncate the half-wave Fourier sine series to $M$ terms in the spatial variable $\mathbf x$ before extending the potential as a Dyson series in the interaction picture to compute the evolution operator matrix elements up to the linear and quadratic integral functionals of $ \mathbf V_n(t)'$s. As a result, we used the Dyson series with the Frobenius norm to reduce the difference between the derived gate energy and the given gate energy, and we determined the temporal performance criterion by plotting the noise-to-signal energy ratio (NSER). A mathematical explanation for a quantum gate's magnetic control has also been provided. In addition, we provide a mathematical explanation for a quantum gate that uses magnetic control.}

\end{abstract}

\section{Introduction}
\label{intro}
\small{Quantum mechanics can be comprehended in a straightforward manner by making use of the atomic interaction picture \cite{Ref1} in the appropriate context. W. Heisenberg and E. Schr$\ddot{o}$dinger both contributed to the development of the theories that underpin quantum mechanics during the first half of the 20th century. This is a fundamental tenet of quantum mechanics and is referred to as a "cornerstone" in the field \cite{Ref1}. The Schr$\ddot{o}$dinger picture allows for a more intuitive understanding of the evolution of a quantum system. In this picture, the wavefunction is the fundamental object and the Hamiltonian is used to describe how the wavefunction evolves. This is in contrast to the Heisenberg picture, where the operators are fundamental objects and the wavefunction is used to describe the evolution of the system.  The Schrodinger picture also has the benefit of being easier to understand than the Heisenberg picture. This is used in a variety of quantum computing applications, such as quantum teleportation, quantum error correction, and quantum search algorithms, which can be used to develop efficient algorithms for simulating the evolution of quantum systems and can provide insight into the behavior of quantum systems.
	
Quantum physics is helping to facilitate the development of the next generation of computers, which will have performance capabilities that are superior to those of today's computer technology. Instead of bits, however, qubits are used to store the information. It is possible to express different states of qubits as complex linear superpositions of bits. Hadamard, Pauli, and controlled unitary gates are just a few of the many types of gates that can be utilized in the synthesis of the multi-level logic circuits that are referenced in \cite{Ref8}-\cite{Ref12}. It is possible to locate the quantum state in a linear vector space \cite{Ref5}-\cite{Ref7}. Due to the fact that all quantum transformations are unitary and capable of being reversed, the quantum gates themselves also exhibit this property in nature. As a consequence of this, developing quantum algorithms is extremely difficult because they have to be reversible in order to work properly \cite{Ref13}. This means that instead of the simple application of traditional logic circuits, a set of carefully designed quantum algorithms is required for the precise manipulation and transformation of qubits. The use of quantum gates, combined with quantum algorithms and precise qubit manipulation, is necessary for the development of a high-level logic circuit capable of performing any desired function. This concept is fundamental in quantum mechanics, as it allows for the prediction of measurement outcomes and the calculation of probabilities. Observables are also closely related to the concept of eigenvalues and eigenvectors, which play a crucial role in understanding the behavior of quantum systems. An observable can be defined as the linear superposition of a group of orthogonal projections using the spectral theorem for Hermitian operators. Because of this, each observable can be associated with a distinct color spectrum (PVM). In a more general sense, POVMs can be understood as the transformation of a full spectral family into a set of identity-conserving positive operators. Alternatively, the term "generalized measurements" can be used to describe these standards. 
	
In this paper, we demonstrate quantum gates using a small electric field perturbation of a free particle in a box \cite{Ref14}. The problem is discussed in order to be used in the design of various quantum gates, which aid in the generation of quantum logic circuits that are faster than conventional logic circuits. To realize the quantum gates, we perturbed the Hamiltonian with a small electric field, as previously discussed in \cite{Ref15}-\cite{Ref18}. When a quantum system is perturbed by electric, magnetic, or electromagnetic radiation, it becomes excited and changes its state. The Hamiltonian can be perturbed with an electric field by perturbing the free Hamiltonian $H_0$ with a potential $V$ in the physical theories. Quantum mechanical principles allow for practical application of the physical theories cited in \cite{Ref2}. and the eigenvalues of matrices of that theory can be used to represent various observables, including position, momentum, and energy, and the eigenvalues give the results of these observables \cite{Ref2}. Because Hermitian matrix eigenvalues are generally accepted to be real, Hermitian matrices are used as a representation for these quantities of interest. This is because, according to quantum mechanics, the eigenvalues of an observable represent the set of all possible outcomes of a measurement of that observable. This is due to the fact that there must be a real world existence of any measurable quantity.
\subsection {Advantage of perturbation theory}
The advantage of perturbation theory is that it can increase the dimensionality of the matrices in operator space to infinity and find an approximate solution to the corresponding Schr$\ddot{o}$dinger equation that best approximates the given unitary gate. The perturbing time-dependent potential is chosen in such a way that the unitary evolution operator at time $T$ up to second-order becomes as close to the desired unitary gate on the same Hilbert space as possible. The perturbed Hamiltonian is made up of a free particle and a perturbed potential $ V(t,x)$, where $ -\frac{\partial V(t,x)}{\partial x}=E(t,x)$ is the applied electric field. We design the control potential $V(t,x) $ so that the unitary evolution operator of the unperturbed system, after truncation to a finite number N of energy levels, approximates a given unitary gate $U g$ of size $ N\times N $, such as the quantum Fourier transform gate \cite{Ref20},\cite{Ref21}. The potential is chosen in such a way that the desired gate error energy is minimized, which is equal to the Frobenius norm square of the error between the given gate and the Dyson series-based evolved gate. The gate error energy is expanded in the potential up to the quadratic terms, so that the optimal error energy minimization equations are linear integral equations for the potential over $ t\epsilon[0, T] $. The discretization approach in MATLAB is used to solve these coupled linear integral equations, which tend to replace linear integral equations with linear matrix equations that are simple to invert \cite{Ref22}.
\subsection {Dyson series unitary preservation}
The Dyson series has proven to be a useful tool in the study of quantum systems, and it has also been applied successfully to the field of quantum computing. In particular, the Dyson series can be employed in the determination of energy levels and the investigation of quantum system dynamics. Also, entanglement and decoherence effects can be investigated with this method. Quantum computing researchers have used the Dyson series to create new algorithms and analyze the impact of noise on the field. In addition, researchers have used the Dyson series to learn more about quantum error correction and create cutting-edge quantum computing methods. Dyson series quantum unitary gates are novel in that they can be used to investigate the impact of background noise on quantum computing and to design improved algorithms for the field. Decoherence and entanglement in a system are two other topics that can be explored with the Dyson series. The Dyson series can also be used to investigate the consequences of quantum error correction and to create novel quantum computing methods. To investigate how quantum unitary gates react to disturbances, the Dyson series can be used. Both the external environment and the system's own dynamics can contribute to perturbations. New algorithms for quantum computing can be developed and the accuracy of existing algorithms improved by studying the effects of perturbations on quantum unitary gates \cite{Ref23}-\cite{Ref24}. The Dyson series can also be used to learn more about decoherence and create better quantum computing methods. As a conclusion, the Dyson series is an effective method for researching quantum systems, and it has significant applications in quantum computing. Quantum energy levels and system dynamics can be investigated using this method. As an added bonus, it can be used to investigate system entanglement and the consequences of decoherence. In addition, new quantum computing algorithms can be developed by analyzing the effects of perturbations on quantum unitary gates using the Dyson series. By utilizing the Dyson series, we can effectively study quantum systems and use it to develop new algorithms for quantum computing \cite{Ref25}.
\subsection {Our work's novelty}
In several ways, the work is novel: Initially, we construct a quantum unitary gate utilizing the most fundamental physical system in quantum mechanics with discrete energy levels, a free particle in a box. The Hamiltonian perturbation is the most general type of perturbation, consisting of a non-uniform electric field in space and time. As a result, we are optimizing the error energy over a wide range of perturbations. The novel's concluding characteristic involves the creation of massive gates. Increasing the number of base energy states allows us to create unitary gates of nearly infinite size. The primary reason is that there is no such thing as a 1-D or 2-D box; only 3-D boxes exist. We confine an ion in a three-dimensional box, excite it with an electric field that varies along the X-axis only, and then apply our time-dependent perturbation theory to it. Recent developments in quantum computing have enabled the realization of quantum gates using small perturbations of a free particle in an electric-field-enclosed box. This technique has been used to create a quantum gate, which is a device that can alter the state of a quantum system. This article explores the theoretical and experimental aspects of this technique, as well as its potential applications. Using a small perturbation of a free particle in a box containing electric fields to create a quantum gate is the concept underlying this technique. This is accomplished by applying a small electric field to the box, causing the particle to move in a particular direction. This motion can be used to manipulate the state of the particle, thereby enabling the creation of a quantum gate. Understanding the dynamics of the particle in the box and the effects of the electric field on the particle is required for the theoretical aspects of this technique. This requires a comprehensive evaluation of the particle's motion and the electric field's effect on the particle. When designing the quantum gate, it is also necessary to consider the effects of the electric field on the particle \cite{Ref26}. 

Quantum computing relies heavily on the Frobenius norm and the fidelity metric to measure the size and precision of quantum circuits. This is a crucial measurement technique. These metrics can be used to compare various quantum circuits and determine which are more efficient and precise. In the future, these metrics may be employed to enhance the performance of quantum computers and quantum circuits. In addition, these metrics could be utilized to compare quantum algorithms and determine which ones are the most effective and precise. This could facilitate the improvement of quantum computing algorithm development. Lastly, these metrics could be used to compare quantum technologies and determine which is the most dependable and effective. Creating the electric field and measuring the motion of the particle are the experimental aspects of this method. This requires specialized equipment, such as lasers and detectors, to measure the particle's motion and the electric field's effect on the particle. In addition, the electric field must be precisely regulated so that the particle's motion can be precisely measured. Among the potential applications of this method are the development of quantum gates for quantum computing, quantum cryptography, and quantum sensing. This technique could also be used to produce quantum gates, which could be utilized for quantum teleportation and quantum communication. The realization of quantum gates using small perturbations of a free particle in a box with electric fields is a promising technique with the potential to revolutionize quantum computing. This technique has the potential to facilitate the creation of quantum gates for applications such as quantum computing, quantum cryptography, quantum sensing, quantum teleportation, and quantum communications. Further investigation is required to fully comprehend the theoretical and experimental aspects of this technique and to investigate its potential applications. Theoretically, this technique would allow the production of quantum gates with small perturbations of a free particle in an electric field, creating a great potential for numerous applications \cite{Ref27}.

 The magnetic field-based gate design is more complicated here because a 1-D magnetic field will not act on a 1-D charged particle $(q\vec{V}\times \vec{B}\ is \perp \vec{V})$. So we used to excite a charged particle in a 3-D box with a 3-D control magnetic field and then design the gate. Consequently, the magnetic field must be expressed as
 
 $$B(t, X,Y,Z)=\sum_{nmr} B_{nmr}(t)\sin\bigg(\frac{n\pi X}{a}\bigg)\sin\bigg(\frac{n\pi Y}{b}\bigg)\sin\bigg(\frac{n\pi Z}{c}\bigg)$$
 and $\{B_{nmr}(t)\}$ must be optimally determined in control with electric field exactly $\vec{V}(t,X)=\sum_n\vec{V}_n(t)\sin\bigg(\frac{n\pi X}{a}\bigg)$. The magnetic field is expressed with both orbital and spin angular momentum. 
 
 $$H_I(t)=\frac{e(\vec{B}(t,X,Y,Z),\vec{\sigma})}{2m}+\frac{e(\vec{B}(t,X,Y,Z),\vec{L})}{2m}$$
 Where, $L=-\iota\vec{r}\times \vec{\bigtriangledown}$. $B_{nmr}(t)$ depends on $3$ space Fourier indices $(nmr)$ in control with $V_n(t)$ which  depends on only one index. Therefore, magnetic field-based gate design is more difficult to compute.
 
 The paper is structured as follows: First, we compute the first-order perturbation for the Schr$\ddot{o}$dinger evolution operator when the Hamiltonian of a particle in a box is perturbed by a small time-varying potential. Then, in terms of minimizing the Frobenius norm, we define the minimum error energy between the perturbed generator and the given generator. We argue that using the time-dependent perturbation theory of independent quantum systems in conjunction with the matching generator technique is a natural way to realize non-separable gates, which are small perturbations of separable gates. The optimal design's obtained integral equation is then discretized to yield a recursive algorithm. In Section 3, we also evaluate the performance of our algorithm by using the error energy square in the generator as noise power and the given generator energy square as signal power, which is referred to as the noise-to-signal energy ratio (NSER). In Section 4, we justify the use of magnetic field control in the design of a gate. Section 5 concludes with a discussion of future scope.}

\section{Mathematical Model of Quantum Unitary Gate}

The Hamiltonian of a free particle in a 1-D box $[0, L]$ is $H_0 = -\frac{1}{2}\frac{d^2}{dx^2}$. The boundary conditions for the wave function are $\Psi(0,t) = \Psi(L,t) = 0$. The stationary wave functions are obtained by solving the stationary (time-independent) Schr$\ddot{o}$dinger equation $-\frac{1}{2}u{''}(x) = Eu(x)$ with the boundary conditions $ u(0) = u(L) = 0$. The solutions are provided by \cite{Ref3}:
$$u(x) = c_1 \sin(\sqrt{2E}x) + c_2\cos(\sqrt{2E}x)$$ Boundary conditions give $c_2 = 0$,
$$L\sqrt{2E} = n\pi, n = 1, 2, 3....$$
So $u_n(x) = c_n \sin (\frac{n\pi x}{L}).$
On normalization,we get: $\int
_0^L u_n^2(x)dx = 1$ so that $Lc_n^2 = 2$, or, $c_n = \sqrt{\frac{2}{L}}$ and $E_n = \frac{n^2\pi^2}{2L^2}$. A linear superposition of the stationary states yields the general quantum state, which adheres to the normalization constraint:  
$$\Psi(x,t) = \sum_{n=1}^\infty c_n e^{-\iota E_n t}u_n(x)$$
$$\sum_{n=1}^\infty |c_n|^2 = 1.$$
We apply a small perturbation $\epsilon V(t,x)$ so the perturbed Hamiltonian is
$H(t) = -\frac{1}{2}\frac{d^2}{dx^2} + \epsilon V(t,x)$. The Schr$\ddot{o}$dinger's equation for the perturbed hamiltonian becomes
\begin{equation*}
i\frac{d\Psi(x,t)}{dt} = (H_0 + \epsilon V(t))\Psi(x,t).
\end{equation*}
According to this notation, the multiplication operator $V(t)$ multiplies the state by $V(t,x)$. The following characteristics of the evolution operator $U_-t$ can be used to describe it:
\begin{equation*}
\Psi(x,t) = \Psi_t(x) = U_t \Psi_0(x).
\end{equation*}
\begin{equation*}
i\frac{dU_t}{dt} = H(t)U_t = (H_0 + \epsilon V(t))U_t,
\end{equation*}
We wish to describe this unitary evolution entirely in terms if the interaction potential $\epsilon V\left(t\right)$ only, thereby removing the unperturbed Hamiltonian $H_{0}$. This is achieved via the interaction picture representation of Dirac in which observables evolve according to the unperturbed Hamiltonian $H_{0}$, while the states evolve according to the "rotated perturbation" $\tilde{V}\left(t\right) = e^{itH_{0}}V\left(t\right)e^{-itH_{0}}$. To see how this is done, we set $U_t = e^{-itH_0}W_t$ and obtain $i\frac{dW_t}{dt} = \epsilon \widetilde{V_t}W_t.$ \cite{Ref1},\cite{Ref2}.
Now $\widetilde{V_t} = e^{itH_0}V_t e^{-itH_0}.$\\
Suppose, $$ \Psi_0(x) = u_n(x)=|n \rangle$$
then,
\begin{align}
\nonumber U_t(x) =& e^{-iE_nt}u_n(x)-i\epsilon\int_0^te^{-iH_0t}\widetilde{V}_{t_1}u_n(x)dt_1
\\&-\epsilon^2\int_{0<t_2<t_1<t}e^{-i
H_0t}\widetilde{V}_{t_2}\widetilde{V}_{t_1}u_n(x)dt_2dt_1+O(\epsilon^3).
\end{align}
Consider the matrix elements of $U_t$ in the basis of stationary states of the unperturbed Hamiltonian:
\begin{align}
 \nonumber\langle m|U_t|n\rangle& = \langle m|\Psi_t\rangle = \int_0^L u_m(x)\Psi_t(x)dx
= e^{-i E_mt}\delta_{mn}
\\&\nonumber-i\epsilon\int_0^t\langle
m|V_{t_1}|n\rangle e^{-i (t-t_1)E_m}e^{-i t_1E_n}dt_1
\\&-\epsilon^2\int_{0<t_2<t_1<t}\!\!\!\!\!\!\!\!\!\!\!\!\!\!\!\!\!\!\!\!\!\!\!\!e^{-i (t-t_1)E_m}e^{-i
(t_1-t_2)E_p}e^{-i E_nt_2}
\langle m|V_{t_1}|p\rangle \langle
p|V_{t_2}|n\rangle dt_2dt_1+O(\epsilon^3).
\end{align}
Or,
\begin{align}
 \nonumber e^{i E_mt}\langle m|U_t|n\rangle = &\delta_{mn}-i\epsilon\int_0^t\langle m|V_{t_1}|n\rangle
e^{i w_{mn}t_1}dt_1
\\&-\epsilon^2\int_{0<t_2<t_1<t}\!\!\!\!\!\!\!\!\!\!\!\!\!\!\!\!\!\!\!\!\langle m|V_{t_1}|p\rangle \langle
p|V_{t_2}|n\rangle e^{i(w_{mp}t_1+w_{pn}t_2)}
dt_2dt_1+O(\epsilon^3).
\end{align}
Note that, where $w_{mn}$ is a Bohr frequency. In quantum computing is a measure of the energy that is required to perform a quantum computation. This energy is used to manipulate qubits, the fundamental building blocks of quantum computers. In a quantum computer, qubits are manipulated using a variety of techniques, such as entanglement, superposition, and measurement. The Bohr frequency is a measure of the energy required to perform a single operation on a qubit, such as entangling two qubits or measuring the state of a single qubit.  The frequency is expressed in terms of the energy gap between the two energy levels of the qubit, and is typically measured in GHz (gigahertz). The Bohr frequency is important in the development of quantum computers because it allows researchers to accurately measure the energy required to perform a single operation. This knowledge can be used to optimize the performance of the system, as well as to compare the performance of different quantum computers. Additionally, the Bohr frequency can be used to understand the behavior of quantum states, as well as to identify potential errors in the system. Overall, the Bohr frequency is an important concept in quantum computing, as it provides a measure of the energy required to perform a single operation on a qubit. It is an essential aspect of developing and optimizing quantum computers, as well as understanding the behavior of quantum states. 
\begin{align}
\langle m|V_{t_1}|n\rangle = \frac{2}{L}\int_0^L\sin(\frac{m\pi x}{L})\sin(\frac{n\pi
x}{L})V(t,x)dx.
\end{align}
We expand,
\begin{align}
V(t_1,x)= \sum_{n=1}^\infty V_n(t)\sin(\frac{n\pi x}{L}).
\end{align}
If the free particle has charge $Q$ and $V(t)$ is obtained by
applying an electric field $E(t,x)$ Then,$V(t,x)= -Q\int_0^xE(t,\xi)d\xi.$

The electric field $E(t,x)$ may be generated by inserting external
probes at different point $x\in[0, L]$. Each probe has a resistor
and if $R(x)$ is the resistance per unit length then the total power
dissipated in these resistors at time $t$ is proportional to,
\begin{align}
 \nonumber P(t) = \int_0^LR(x)E^2(t,x)dx\\
 \nonumber \propto
\int_0^LR(x)\left(\frac{\partial V(t,x)}{\partial x}\right)^{2}dx.
\end{align}
Expanding $R(x)$ as a half wave Fourier series as $R(x) = \sum_{n=1}^\infty R(n)\sin\left(\frac{n\pi x}{L}\right)$, we
get,\cite{Ref22}R$$P(t)\propto \sum_{n=1}^\infty
\left(\frac{n\pi}{L}\right)^2R_nV_n^2(t).$$ 
Now,
\begin{align}
\langle m|V_t|n\rangle \!\!= \!\!\!\!\sum_k\!\!\frac{2}{L}\!\!\int_0^L\!\!\!\!\!\!\sin\left(\frac{m\pi x}{L}\right)
\sin\left(\frac{n\pi x}{L}\right)V_k(t)\sin\left(\frac{k\pi x}{L}\right)dx.
\end{align}
Therefore, using $$\int_0^L\sin\left(\frac{n\pi x}{L}\right)dx =
\begin{cases}0
  \text{,if $n=0$}\\\dfrac{L}{n \pi}(1-(-1^n)) \text{,if $n\neq 0$}
\end{cases}$$
we get,
\begin{align}
\nonumber\langle m|V_t|n\rangle =&
\sum_k\frac{1}{2\pi}\bigg\{\frac{1-(-1)^{k+m-n}}{k+m-n}+\frac{1-(-1)^{k+n-m}}{k-m+n}
 \\\nonumber &-\frac{1-(-1)^{k+m+n}}{k+m+n}-\frac{1-(-1)^{k-m-n}}{k-m-n}\bigg\}V_k(t).
\end{align}
Finally,
\begin{align}
\nonumber \langle m|V_t|n\rangle =&
\frac{1}{2\pi}\bigg\{\sum_{r\geq(\frac{m-n}{2})}\frac{2}{2r+1}V_{2r+1-m+n}(t)
+\sum_{r\geq(\frac{n-m}{2})}\frac{2}{2r+1}V_{2r+1+m-n}(t)
 \\\nonumber &-\sum_{r\geq(\frac{m+n}{2})}\frac{2}{2r+1}V_{2r+1-m-n}(t)
-\sum_{r\geq(-\frac{m+n}{2})}\frac{2}{2r+1}V_{2r+1+m+n}(t)\bigg\}.
\end{align}
\begin{align}
\langle m|V_t|n\rangle = \sum_{p=1}^\infty k[m,n,p]V_p(t).
\end{align}

Thus, we are truncated by t = T and discretization method have to use for simulation;
\begin{align}
\nonumber e^{i E_m T}&\langle  m|U_T|n\rangle = \delta[m-n]
-i\epsilon\int_0^T k[m,n,p]V_p(t)e^{i w_{mn}t}dt
\\&-\epsilon^2\int_{0<t_2<t_1<T}\!\!\!\!\!\!\!\!\!\!\!\!\!\!\!\!\!\!\!\!\!\!\!\!k[m,n,p]k[p,n,r]
V_q(t_1)V_r(t_2)e^{\iota(w_{mp}t_1+w_{pn}t_2)}dt_2dt_1+ O(\epsilon^3),
\end{align}
where, summation over the repeated indices $p,q$ is implied. Let
$$
U_d[m,n] = \langle m|U_d|n\rangle = U_d[x, y].
$$
\begin{align}
=\frac{2}{L}\int_0^L\int_0^L\sin(\frac{m\pi x}{L})\sin(\frac{n\pi
x}{L})U_d(x,y)dxdy.
\end{align} Then, define
\begin{align}
 W_d[m,n] =
\delta[m-n]-e^{-iE_n T}U_d[m, n].
\end{align}
We have,

$$
H=-\frac{1}{2}\frac{d^2}{dx^2}+\epsilon.V(t,x), 0\leq x\leq L.
$$
Energy levels of the unperturbed system are$E_n=n^2\pi^2/2L^2$
Take $L=1$. Note that $m=1$. Normalized energy eigenstate
corresponding to $E_n$:
$$
\ket{n}=\sqrt 2.sin(n\pi x), n=1,2,...
$$
In quantum computing, the Schr$\ddot{o}$dinger interaction picture is an alternative to the Heisenberg picture and allows for the evolution of a quantum system to be studied in terms of a unitary operator. It is particularly useful for dealing with fast-changing external fields or when certain degrees of freedom in the system are changing rapidly. It is based on the wavefunction of the system and is useful for studying the evolution of quantum states in the presence of a Hamiltonian. In the Schr$\ddot{o}$dinger picture, the Hamiltonian is divided into two parts: one that is time-dependent and one that is fixed. The time-dependent part is used to describe the evolution of the system and the fixed part is used to describe the initial state of the system. The Schr$\ddot{o}$dinger picture is useful for analyzing the effects of a time-dependent Hamiltonian on the system, such as the effects of time-dependent forces and evolution operator satisfies \cite{Ref3}
$$
iU'(t)=(H_0+\epsilon V(t))U(t),\text{where } H_0=-\frac{1}{2}\frac{d^2}{dx^2},
$$
$$
U(t)=exp(-itH_0)W(t).
$$
Then,
$$
iW'(t)=\epsilon.\tilde V(t)W(t).
$$
$$
\tilde V(t)=exp(itH_0)V(t)exp(-itH_0).
$$
We get,
$$
W(T)=I-i\epsilon\int_0^T\!\!\!\!\!\!\tilde
V(t)dt-\epsilon^2\int_{0<t_2<t_1<T}\!\!\!\!\!\!\!\!\!\!\!\!\!\!\!\!\!\!\!\!\!\!\!\!\!\!\!\\\tilde V(t_1)\tilde
V(t_2)dt_2dt_1+O(\epsilon^3).
$$
Let, $U_d$ be the desired unitary gate. We wish to choose $V(t,x)$
over $(t,x)\in[0,T]\times[0,L]$ so that $\parallel
U_d-U(T)\parallel$ is minimized upto $O(\epsilon^2)$ terms and expending through Dyson series. Now,
\begin{align}
\!\!\!\!\!\!\!\!\!\!\!\!\!\!\!\!\!\!\!\!\!\nonumber&\parallel U_d-U(T)\parallel^2=\parallel U_d-exp(-itH_0)W(T)\parallel^2
\\&\nonumber=\parallel exp(itH_0)U_d-I+i\epsilon\int_0^T\!\!\!\!\!\!\tilde V(t)dt
\\&+\epsilon^2\int_{0<t_2<t_1<T}\!\!\!\!\!\!\!\!\!\!\!\!\!\!\!\!\!\!\!\!\!\!\!\!\tilde V(t_1)\tilde
V(t_2)dt_2dt_1\parallel^2+O(\epsilon^3).
\end{align}
where $\parallel(.)\parallel$ stands for the Frobenius norm, defined as:
\begin{equation*}
\parallel X \parallel^2 = Tr\left(X^*X\right)
\end{equation*}
Setting,
$$
W_d=exp(itH_0)U_d-I.
$$
we have,
\begin{align}
\nonumber&\parallel U_d-U(T)\parallel^2=\parallel W_d\parallel^2
+\epsilon^2\int_{0<t_1,t_2<T}Tr(\tilde
V(t_1)\tilde V(t_2))dt_1dt_2
\\&+2\epsilon\int_0^TIm(Tr(W_d\tilde V(t)))dt
+2\epsilon^2\int_{0<t_2<t_1<T}\!\!\!\!\!\!\!\!\!\!\!\!\!\!\!\!\!\!\!\!\!\!\!\!Re(Tr(W_d\tilde V(t_2)\tilde V(t_1)))dt_2dt_1
+O(\epsilon^3).
\end{align}
We expand the external potential,
$$
V(t,x)=\sum_nV_n(t)sin(n\pi x).
$$
The energy constraint is in quantum computing is a powerful tool, but it can be very energy intensive. To make the most of quantum computing, it is important to understand and manage the energy constraints it imposes. One of the most important energy constraints in quantum computing is the power budget. Quantum computers require large amounts of power to run, and they are subject to a “power wall” that limits the amount of power they can consume. This means that the total power consumed by a quantum computer must be kept within certain bounds in order to ensure that the device does not become unstable and fail. Another energy constraint in quantum computing is the cooling requirements. Quantum computers must be kept extremely cold in order to function properly. This means that large amounts of energy must be used to cool the computers and keep them at the required temperature. 
Finally, quantum computers must be carefully shielded from external sources of noise and interference. This requires the use of additional energy, both for the shielding itself and for the cooling required to keep the shielding at the optimum temperature. Overall, quantum computing imposes a number of energy constraints that must be taken into account in order to get the most out of the technology. By understanding these constraints and managing them effectively, quantum computing can become a powerful and efficient tool.
$$
\sum_{p=1}^N\alpha[p]\int_0^TV_p^2(t)dt=E.
$$
This is obtained from,
$$
\int_0^T\int_0^LR(x)V^2(t,x)dtdx=constt,
$$
where, $R(x)$ is a known function (conductance per unit length).
Take $E=E_0/5$, $\alpha[p]=1 (constant)$ for all $p$ and $\epsilon=1$ where,
$E_0=\pi^2/2$ $(\pi^2/2mL^2)$ is the ground state unperturbed
energy. This value of $E$ ensures that the perturbation is indeed
small compared to the unperturbed energy. Further, choose
$T=10/E_0=20/\pi^2$. The ground state evolves as $exp(-iE_0t)$,
ie, with frequency $E_0$ and hence $1/E_0$ is the longest time
scale involved in the unperturbed system. We have chosen $T$ to be
$10$ times this longest time scale to allow sufficiently good gate
approximation. Now,
$$
Tr(W_d\tilde V(t))=\sum_{m,n=1}^NW_d[m,n]\langle n|\tilde V(t)|m\rangle.
$$
\begin{align}
=2\sum_{m,n,p}W_d[m,n]V_p(t)exp(i\omega[nm]t)
\int_0^1sin(n\pi x)sin(m\pi x)sin(p\pi x)dx,
\end{align}
where,
$$
w_{mn}=\omega[nm]=E_n-E_m=(n^2-m^2)\pi^2/2.
$$
Let,
$$
\gamma[mnp]=2\int_0^1sin(n\pi x)sin(m\pi x)sin(p\pi x)dx.
$$
$$
\approx (2/M)\sum_{r=0}^{M-1}sin(n\pi r/M)sin(m\pi r/M)sin(p\pi
r/M),
$$
where, $M=100$.

Thus,
$$
2\epsilon\int_0^TIm(Tr(W_d\tilde V(t)))dt=
$$
$$
4\int_0^T\!\!\sum_{m,n,p=1}^N\!\!\!\!\!\!Im(W_d[m,n]exp(i\omega[nm]t))\gamma[mnp]V_p(t)dt
$$
$$
=\sum_{p=1}^N\int_0^T\beta_p(t)V_p(t)dt,
$$
where,
$$
\beta_p(t)=4\sum_{m,n}\gamma[mnp]Im(W_d[m,n]exp(i\omega[nm]t)).
$$

Note that, $\left(\left(W_d[m,n]\right)\right)$ is an $N\times N$ matrix and so is
$\omega[n,m]=E_n-E_m$. These have to be defined before implementing
the above program. To compute and store $W_d[n,m]=\langle n|W_d|m\rangle$, we
observe that,
$$
W_d=exp(iTH_0)U_d-I,
$$
so that,
$$
W_d[m,n]=exp(iE_mT)\langle m|U_d|n \rangle-\delta[m-n].
$$
$U_d$ will typically be given as a kernel $U_d(x,y)$ from which
$U_d[m,n]$ will be calculated as,
$$
U_d[m,n]=2\int_0^1\int_0^1U_d(x,y)sin(m\pi x)sin(n\pi y)dxdy
$$
$$
\approx (1/M^2)\sum_{r,s=0}^{M-1}\!\!\!\!U_d(r/M,s/M)sin(m\pi r/M)sin(n\pi
s/M).
$$
A program can be written to compute and store $U_d[m,n]$ as an
$N\times N$ matrix. Or, we may take for the purpose of
illustration, $U_d[m,n]=exp(2\pi i(m-1)(n-1)/N)/\sqrt N$ with
$1\leq m,n\leq N$, i.e. the DFT matrix. We may define the vector
$E=\left(\left(E(n)\right)\right)_{n=1}^N$, of unperturbed energies also beforehand
$E(n)=n^2\pi^2/2$. Define the diagonal matrix,
$$
D=diag[exp(iE(m)T), m=1,2,..., N],
$$
and then compute,
$$
W_d=D*U_d-I.
$$
We get the approximation,
$$
2\epsilon\int_0^TIm(Tr(W_d\tilde V(t)))dt
$$
$$
\approx\sum_{p,r}\beta[K(p-1)+r+1]V_p(r\tau)\tau.
$$

(We have not yet obtained $V$, we are proceeding that $V$ has been
stored in this form as an unknown $KN\times 1$ vector)

We have,
\begin{align}
2\epsilon\int_0^TIm(Tr(W_d\tilde
V(t)))dt\approx\sum_{p,r}\beta[K(p-1)+r+1]
V[K(p-1)+r+1]\tau.
\end{align}
$$
=\tau.\sum_{m=1}^{KN}\beta[m]V[m].
$$

Likewise, we compute,
$$
Term(2)=\int_{0<t_1,t_2<T}Tr(\tilde V(t_1)\tilde V(t_2))dt_1dt_2.
$$
\begin{align}
=\int_{0<t_1,t_2<T}\sum_{m,n=1}^Nexp(i\omega[m,n](t_1-t_2))
\langle m|V(t_1)|n\rangle\langle n|V(t_2)|m \rangle dt_1dt_2.
\end{align}
 We now propose to replace all continuous-time integrals by discrete sums over integer time indices with $\tau$ as the time discretization step size. Further, we club the discrete-time index and the potential integer index into a single index by the process of lexicographic ordering. This clubbing enables us to formulate the problem of determining the optimal potential as a matrix-vector combined quadratic optimization problem, which can be optimized using MATLAB \cite{Ref15},\cite{Ref16}.
 Again, we observe that,
 $$
 \langle m|V(t)|n \rangle=\sum_p\gamma[mnp]V_p(t),
 $$
 and hence,
 \begin{align}
 Term(2)=\sum_{mnpq=1}^N\gamma[mnp]\gamma[nmq]\int
 V_p(t_1)V_q(t_2)
 cos(\omega[m,n](t_1-t_2))dt_1dt_2.
 \end{align}
 \begin{align}
 =\tau^2\!\!\!\!\!\!\!\!\sum_{mnpq,r_1r_2}\!\!\!\!\!\!\!\!\gamma[mnp]\gamma[mnq]V_p(r_1\tau)V_q(r_2\tau)
 cos(\omega[m,n](r_1-r_2)\tau).
 \end{align}
 \begin{align}
 \nonumber=&\tau^2\sum\gamma[mnp]\gamma[mnq]cos(\omega[m,n](r_1-r_2)\tau)
 \\& \times V[K(p-1)+r_1+1]V[K(q-1)+r_2+1].
 \end{align}
 Thus, defining the $KN\times KN$ matrix $C$ with entries,
 $$
 C[K(p-1)+r_1+1,K(q-1)+r_2+1]
 $$
 $$
 =\tau^2\sum_{m,n=1}^N\gamma[mnp]\gamma[mnq]cos(\omega[m,n](r_1-r_2)\tau).
 $$
 for $1\leq p,q\leq N$, $0\leq r_1,r_2\leq K-1$, we have,
 $$
 Term(2)=
 $$
 $$
 =\sum_{m,n=1}^{KN}C[m,n]V[m]V[n].
 $$
 Likewise,
 $$
 Term(3)=\int_{0<t_2<t_1<T}\!\!\!\!\!\!\!\!\!\!\!\!\!\!\!\!\!\!\!\!\!\!\!\!Re(Tr(W_d\tilde V(t_2)\tilde
 V(t_1)))dt_2dt_1.
 $$
 \begin{align}
 =\int_{0<t_2<t_1<T}\sum_{mnp}Re(W_d[m,n]<n|V(t_2)|p>
 <p|V(t_1)|m>)dt_2dt_1.
 \end{align}
 \begin{align}
 \nonumber=\int_{t_2<t_1}\sum_{mnp}&Re(W_d[m,n]\gamma[npa]\gamma[pmb]
 \\& \times exp(i(\omega[p,m]t_1+\omega[n,p]t_2))V_a(t_1)V_b(t_2)dt_2dt_1.
 \end{align}
 Define the $KN\times KN$ matrix $D$ by,
 \begin{align}
 \nonumber &D[K(a-1)+r_1+1,K(b-1)+r_2+1]=
 \\& =\!\!\!\!\!\!\sum_{mnp=1}^N\!\!\!\!\!\!\gamma[npa]\gamma[pmb] Re(W_d[m,n]
 \times exp(i(\omega[p,m]r_1+\omega[n,p]r_2)\tau)).
 \end{align}
 Note that, $\gamma[mnp]$ is accessed as $\gamma[N^2(m-1)+N(n-1)+p]$
 from the memory. Then,
 \begin{align}
 \nonumber Term(3)=\!\!\!\!\!\!\!\!\!\!\!\!\!\sum_{a,b,r_1,r_2,r_2<r_1}\!\!\!\!\!\!\!\!\!\!\!\!\!&D[K(a-1)+r_1+1,K(b-1)+r_2+1]
 \\& \times V[K(a-1)+r_1+1]V[K(b-1)+r_2+1].
 \end{align}
 This, like $Term(2)$ is a quadratic form in $V$ and can be
 expressed as $V^THV$ where $H$ is derived from $D$ in an obvious
 way. Finally, combining all this, the optimization problem can be
 cast as,
 $$
 \mathrm{minimize} \hspace{1mm} F({\bf V},\lambda)={\bf V}^T{\bf Q}{\bf V}+\beta^T{\bf
 	V}+\lambda({\bf V}^T{\bf R}{\bf V}-E).
 $$
 where, ${\bf Q,R}$ are positive definite matrices and $E$ is a
 given real positive scalar, $\lambda$ is a real scalar, $\beta$ is
 a real vector.\cite{Ref17},\cite{Ref18}
 $$
 \nabla_{\bf V}F=0, dF/d\lambda=0,
 $$
 give,
 $$
 {\bf QV}+\beta+\lambda{\bf RV}=0, {\bf V}^T{\bf RV}=E.
 $$
 The first gives,
 $$
 {\bf V}=-({\bf Q}+\lambda{\bf R})^{-1}\beta,
 $$
 and substituting this into the second gives,
 $$
 \beta^T({\bf Q}+\lambda{\bf R})^{-T}{\bf R}({\bf Q}+\lambda{\bf R})^{-1}\beta=E.
 $$
 This equation is very hard to solve for $\lambda$. Thus, a
 recursive algorithm for calculating ${\bf V}$ would be as follows:
 Start with an initial guess $\lambda^{(0)}$ for $\lambda$. For
 $n=0,1,2,...$ compute,
 $$
 {\bf V}^{(n+1)}=-({\bf Q}+\lambda^{(n)}{\bf R})^{-1}\beta,
 $$
 $$
 \lambda^{(n+1)}=-[{\bf V}^{(n+1)T}({\bf QV}^{(n+1)}+\beta)]/[{\bf
 	V}^{(n+1)T}{\bf RV}^{(n+1)}].
 $$
 
 \section{Simulation Results}
 It shows that with increasing time the NSER decreases. Thus, we can achieve very low NSER's by increasing the duration of the evolution of time \cite{Ref23}.
 %
 
%
 
It represents the noise-to-signal $(NSER)$ versus time plot which shows that NSER decreases with time.
 $$
 NSER=\frac{\|U_d-U(T)\|^2}{\|U_d\|^2}
 $$
 where $U_d$ is the desired gate and $U(T)$ is the simulated gate. The graph shows that the NSER decreases with time and reaches a steady value and cannot be zero as the truncated Dyson series cannot achieve a steady unitary gate. 
 

 We can also design the quantum gate in the presence of noise and extend the potential upto $ \mathcal{O}(\epsilon^k)$. In this article, 

\section{Design a Gate using Magnetic Field Control}
If a particle in a 3-D box is connected with an electric field applied along only the X-direction, then the total Hamiltonian will be 
$$H(t)=-\frac{1}{2m}\bigg(\frac{\partial ^2}{\partial X^2}+\frac{\partial ^2}{\partial Y^2}+\frac{\partial ^2}{\partial Z^2}\bigg)+\epsilon\sum_{n=1} V_n(t)\sin\bigg(\frac{n\pi X}{a}\bigg)$$ The egigenstaes of the unperturbed Hamitonian obtained after applying  boundary conditions are
$$\big |nmp\rangle = \bigg(\frac{2}{L}\bigg)^{\frac{3}{2}}\sin\bigg(\frac{n\pi X}{L}\bigg)\sin\bigg(\frac{n\pi X}{L}\bigg)\sin\bigg(\frac{n\pi X}{L}\bigg)$$ 
and hence in the intraction picture upto $O(\epsilon^2)$, the matrix elements of the evolution operators are
\begin{align}\nonumber\bra{n'm'p'}W(T)\ket{ nmp}=&\delta[n'-n]\delta[m'-m]\delta[p'-p]\\\nonumber&-\iota\epsilon\int_{0}^{T}\bra{n'm'p'}V(t, X)\ket{ nmp}\exp(\iota(E_{n'm'p'}-E_{nmp})t)dt\\\nonumber&-\epsilon^2\sum_{n"m"p"}\int_{0<t_2<t_1<T}\bra{n'm'p'}V(t_1,X)\ket{n"m"p"}bra{n"m"p"}V(t_2,X)\ket{nmp}\\\nonumber&\exp(\iota((E_{n'm'p'}-E_{n"m"p"})t_1+(E_{n"m"p"}-E_{nmp})t_2))dt_1dt_2\end{align}
where, $E_{nmp}=\frac{ \pi^2}{2m_0L^2}(n^2+m^2+p^2)$. It is easy to see that this matrix factorizes into tensor product form:
\begin{align}\nonumber\bra{n'm'p'}W(T)\ket{ nmp}=\bra{n'}W_X(T)\ket{ n}\delta[m'-m]\delta[p'-p]\end{align}
where,
\begin{align}\nonumber\bra{n'}W_X(T)\ket{n}=&\delta[n'-n]\\\nonumber&-\iota\epsilon\int_{0}^{T}\bra{n'}V(t, X)\ket{n}\exp(\iota(E_{n'}-E_{n})t)dt\\\nonumber&-\epsilon^2\sum_{n"}\int_{0<t_2<t_1<T}\bra{n'}V(t_1, X)\ket{ n"}\bra{n"}V(t_2, X)\ket{ n}\\\nonumber&\exp(\iota((E_{n'}-E_{n"})t_1+(E_{n"}-E_{n})t_2))dt_1dt_2\end{align}
Thus our designed gate has the form 
$$W_X(T)\otimes I_N\otimes I_N$$
and hence if $W_g$ is the derived gate, we have to choose $\{V_n(t)\}$ so that $\|W_X(T)\otimes I_N\otimes I_N-W_g(T)\|^2$ is a minimum, that is, minimize
\begin{align}\nonumber& N^2\|W_X(T)\|^2-2Re (T_r((W_X(T)\otimes I_N\otimes I_N)W_g(T)^*)\\\nonumber &=N^2\|W_X(T)\|^2-2Re (T_r W_X(T) (T_{r_{23}}W_g(T)^*))\end{align}
So that design amounts to the same optimization algorithm but with the given state $W_g(T)$ replaced by its particle trace over the Y and Z dimensions $W_g(T)\rightarrow T_{r_{23}}(W_g(T))$.
or equivalently,
\begin{align}\nonumber\bra{nmps}U(T)\ket{n'm'p's'}=\delta_{ss'}\delta[n-n']\delta[m-m']\delta[p-p']\exp(-\iota E_{nmps}T)\end{align}
where, $E_{nmps}=\frac{\pi^2}{2m_0L^2}(n^2+m^2+p^2)+\frac{esB_0}{2m}$, if $B_0$ is a constant then $\bra{ s'}\sigma \ket{ s}=\delta_{s's}$. In this case, after a constant magnetic field perturbation, the evolution operator remains diagonal and hence is not of much varies. Even when the magnetic field depends only on time and not on space, our exact perturbation unitary evolution operator remains diagonal.

\section{Conclusion}
We have minimized discrepancies between the given unitary gate and the gate designed in the presence of a free particle in a 1-D box that is bounded between two walls with infinite potential in the presence of a weak electric field. We have used the perturbation approach to design the gate, which is more appropriate for the physical system such as atoms, molecules, etc. We have used the Dyson series in the interaction picture to calculate the evolution operator matrix elements upto the linear and quadratic integral function of the potential.

We have reduced the disparity between the supplied unitary gate and the gate developed in the presence of a free particle in a 1-D box that is limited by two walls and has infinite potential in the presence of a weak electric field. The gate that we designed is more suitable for physical systems like atoms, molecules, and so on because we employed the perturbation approach. The Dyson series was used to figure out the evolution operator matrix elements up to a linear and quadratic integral function of the potential in the interaction picture. 

In this paper, we have made an attempt to synthesize the quantum Fourier transform unitary gate by perturbing a particle in a box with a non-uniform time-varying electric field. The particle is assumed to carry a charge, and the total Hamiltonian of the particle is given by
$$
H=-\frac{1}{2}\frac{d^2}{dx^2}-e V(t,x), 0\leq x\leq L
$$
with $V(t,x)$ is the perturbing potential corresponding to the non-uniform electric field. The boundary conditions are that the wave function which satisfies the Schr$\ddot{o}$dinger equation vanishes at $x=0$ and $x=L$. The resulting unitary evolution operator is developed into a Dyson series retaining only upto $\mathcal{O}(e)$ terms and this evolution operator is expressed relative to the eigenbasis of the unperturbed system ie, of a free particle which has eigenfunctions
$$
u_n(x)=(2/L)^{1/2}sin(n\pi x/L), n =1,2,...
$$
with corresponding energy eigenvalues being $n^2\pi^2/2mL^2, n=1,2,...$. After truncating this basis, we calculate the $O(e)$ term in the Dyson series expansion by representing $V(t,x)=\sum_nV_n(t)u_n(x)$ with the control functions $V_n(t)$
calculated so that the Frobenius norm error square between the given QFT gate and the realized gate is as small as possible. The optimal equations are coupled linear integral equations for the unknown functions $V_n(t), 0\leq t\leq T, n\geq 1$ and are solved using MATLAB by discretizing linear integral equations into linear matrix equations.

\section{Future Work}
 Suppose that we have designed our quantum unitary gate and obtained a given set of potential coefficients $V_n(t)$. Now suppose, noise enters our system, ie, instead of $V_n(t)$, we have $V_n(t)+w_n(t)$, or equivalently, $V(t,x)$ is replaced by the randomly perturbed potential
 $$
 \sum_n(V_n(t)+w_n(t))u_n(x)
 $$
 Then, we can calculate the change in the evolution operator upto linear orders in the noise processes $w_n(t)$. Thus, the perturbed unitary gate will have the form
 $$
 U(T)=U_0(T)+\sum_n\int_0^Tw_n(t)W_n(t)
 $$
 where the $W_n(t)'s$ are non-random matrices. We can calculate the increase in the mean square gate energy error due to noise.
 
 $$
 \Bbb E(\parallel U_g-U(T)\parallel^2)
 $$
 \begin{align}
 \nonumber =\parallel U_g-U_0(T)\parallel^2+\sum_{n,m}\int_0^T\int_0^T\Bbb
 E(w_n(t)w_m(s))
 \\\nonumber \times Tr(W_n(t)W_n(s))dtds
 \end{align}
 
 Another possible extension to this problem is to consider a particle in a $3-D$ box perturbed by a vector electric field and a vector magnetic field. The Hamiltonian is then:
 \begin{align*}
 H(t)& = \frac{{\left( -i \hbar \nabla + e A(t,r) \right)}^2}{2m} - e V(t,r)
 \\& = H_0 + e V_1 + e^2 V_2
 \end{align*}
 where,
 \begin{equation*}
 H_0 = -\frac{{\hbar }^2 {\nabla}^2}{2m}
 \end{equation*}
 \begin{equation*}
 V_1(t) = -\frac{i \hbar}{2m} \left[ A(t,r) \mathbf{,} \nabla \right] - \frac{i \hbar}{2m} \nabla \cdot A(t,r) - V(t,r)
 \end{equation*}
 \begin{equation*}
 V_2 (t) = \frac{A^2 (t,r)}{2m}
 \end{equation*}
 
 Due to this perturbation, the unitary evolution $U(t)$ must be expanded upto $\mathcal{O}(e^2)$ and then, matching with the desired unitary gate $U_g$ has to be carried out. Specifically,
 
 \begin{equation*}
 U(t) = U_0(t) \left( I + e W_1 (t) + e^2 W_2 (t) \right)
 \end{equation*}
 where $U_0(t) = exp\left( \frac{-i t H_0}{\hbar} \right)$. So, equating each power of $e$ gives
 \begin{equation*}
 i \hbar \frac{d W_1(t)}{dt} = \widetilde{V_1}(t)
 \end{equation*}
 \begin{equation*}
 i \hbar \frac{d W_2(t)}{dt} = \widetilde{V_1}(t) W_1(t) + \widetilde{V_2}(t) 
 \end{equation*}
 where 
 \begin{equation*}
 \widetilde{V_k}(t) = U_0(-t) V_k (t) U_0(t)
 \end{equation*}
 Optimization with respect to $A(t,r)$ and $V(t,r)$ must be carried out so that the gate error energy:
 \begin{equation*}
 \left \lvert \left \lvert U_g - I - eW_1(T) - e^2 W_2(T) \right \rvert \right \rvert^{2}
 \end{equation*}
 is a minimum for the subject of a future prospect. In addition, we can investigate more complex quantum fourier gates and their applications in quantum computing. We can also explore the use of quantum fourier gates in other areas  such as quantum cryptography and quantum communication. Investigating these complex quantum fourier gates and their applications has the potential to revolutionize our approach to problem-solving, allowing us to create more efficient algorithms.
 

 \section*{Data availability statement }
 There are no associated data with my manuscript.
 \section*{Conflicts of interests}
 On behalf of all authors, the corresponding author states that there is no conflict of interest.

\end{document}